\documentclass[%
 aip,
 amsmath,amssymb,
 preprint,%
]{revtex4-1}

\usepackage{graphicx}
\usepackage{dcolumn}
\usepackage{bm}

\usepackage[utf8]{inputenc}
\usepackage[T1]{fontenc}
\usepackage{mathptmx}
\usepackage{etoolbox}

\makeatletter
\def\@email#1#2{%
 \endgroup
 \patchcmd{\titleblock@produce}
  {\frontmatter@RRAPformat}
  {\frontmatter@RRAPformat{\produce@RRAP{*#1\href{mailto:#2}{#2}}}\frontmatter@RRAPformat}
  {}{}
}%
\makeatother

\begin{document}

\preprint{AIP/123-QED}

\title{A Materials Map Integrating Experimental and Computational Data via Graph-Based Machine Learning for Enhanced Materials Discovery}

\author{Y. Hashimoto}
\email{yusuke.hashimoto.b8@tohoku.ac.jp}
\affiliation{ 
Frontier research Institute for Interdisciplinary Sciences, Tohoku University, Sendai, Japan, 980-8577
}
\author{X. Jia}
\affiliation{ 
Advanced Institute for Materials Research (WPI-AIMR), Tohoku University, Sendai, Japan, 980-8577}
\author{H. Li}
\affiliation{ 
Advanced Institute for Materials Research (WPI-AIMR), Tohoku University, Sendai, Japan, 980-8577}
\author{T. Tomai}
\affiliation{ 
Frontier research Institute for Interdisciplinary Sciences, Tohoku University, Sendai, Japan, 980-8577
}



\begin{abstract}
Materials informatics (MI), emerging from the integration of materials science and data science, is expected to significantly accelerate material development and discovery. The data used in MI are derived from both computational and experimental studies; however, their integration remains challenging. In our previous study, we reported the integration of these datasets by applying a machine learning model that is trained on the experimental dataset to the compositional data stored in the computational database. In this study, we use the obtained datasets to construct materials maps, which visualize the relationships between material properties and structural features, aiming to support experimental researchers. The materials map is constructed using the MatDeepLearn (MDL) framework, which implements materials property prediction using graph-based representations of material structure and deep learning modeling. Through statistical analysis, we find that the MDL framework using the message passing neural network (MPNN) architecture efficiently extracts features reflecting the structural complexity of materials. Moreover, we find that this advantage does not necessarily translate into improved accuracy in the prediction of material properties. We attribute this unexpected outcome to the high learning performance inherent in MPNN, which can contribute to the structuring of data points within the materials map.
\end{abstract}

\maketitle


\section{Introduction}

Materials informatics has emerged as a transformative approach to accelerate materials development by integrating data science with materials science~\cite{Rajan2005-dc,Rajan2015-bj,Agrawal2016-dk,Schleder2019-nd,Rickman2019-de}. By applying data science and machine learning techniques, researchers can efficiently identify and design materials with desired properties, significantly reducing the time and costs associated with traditional experimental methods~\cite{Nakayama2023-pa,Iwasaki2022-ne,Chiba2023-zp,Pedersen2021-lb,Motojima2023-bh,Ozaki2020-mc,Lai2023-yu}. The success of such approaches relies heavily on the availability of high-quality, large-scale datasets. 
There are two distinct materials databases: one for computational data and another for experimental data. 
Initiated in 2011, the Materials Genome Initiative has driven the development of computational databases, such as the Materials Project (MP)~\cite{Jain2013-vc} and AFLOW~\cite{Curtarolo2012-qq}, which have been instrumental in systematically collecting and organizing results from first-principles calculations conducted globally.
When combined with advanced data science techniques, these databases enable the development of machine learning models capable of predicting material properties with remarkable accuracy~\cite{Fung2021-mg}.
A critical challenge persists in bridging the gap between theoretical predictions and practical applications.

Materials are generally described by their atomic composition.
However, their properties fundamentally depend not only on their composition but also on structural arrangements.
To address the limitations of simple chemical formulas in capturing structural nuances, researchers have developed graph-based approaches.
The Crystal Graph Convolutional Neural Network (CGCNN) models materials as graphs, where nodes correspond to atoms and edges represent interactions~\cite{Xie2018-rf}.
This method encodes structural information into high-dimensional feature vectors that, when combined with deep learning techniques, enable the development of robust property prediction models~\cite{Fung2021-mg}.

The application of graph-based representation of materials properties relies heavily on structural data availability~\cite{Xie2018-rf}.
While computational databases provide extensive datasets derived from first-principles calculations, experimental data remains sparse, inconsistent, and often lacks the structural information necessary for advanced modeling.
Although integrating these databases could reveal correlations between experimental data and structural information, direct integration remains challenging due to discrepancies in composition and data format.
This limitation poses challenges in applying graph-based methods to experimental data, leading most studies to rely primarily on computational datasets for such representations~\cite{Fung2021-mg}.

StarryData2 (SD2) is an innovative database project in the field of materials science~\cite{Katsura2019-ox}. Developed by Katsura $et$ $al$., this database aims to systematically collect, organize, and publish experimental data on materials from previously published papers. SD2 has extracted information from over 7,000 papers for more than 40,000 samples. The database covers a wide range of material fields, including thermoelectric materials, magnetic materials, thermal conductivity materials, and piezoelectric materials. Designed for open access, SD2 allows researchers and engineers to freely access and utilize the data.


In our previous study~\cite{Jia2024-qq}, we proposed a novel framework to integrate experimental and computational datasets to create a comprehensive dataset for materials informatics. 
This dataset comprises over 1,000 materials, each described by its structural information and predicted experimental $zT$ values.
The structural information includes atomic positions and lattice parameters, enabling the construction of graph-based representation of materials by MDL~\cite{Fung2021-mg}, which provides a flexible environment for material property prediction.

In this study, we use this dataset to construct material maps.
The creation of interpretable material maps that reflect real-world variability provides a novel framework for efficient material discovery.
The contributions of this study are summarized as follows:

\begin{itemize}
\item We construct materials maps reflecting the relation between the structural complexity and the experimental thermoelectric properties ($zT$ values).
\item The effectiveness of the MPNN architecture in capturing structural complexity for the material map construction is demonstrated.
\item Obtained materials maps are interpreted by making plots colored with various properties stored in a computational database.
\item We highlight the potential of material maps as a tool for guiding experimentalists in synthesizing new materials, enabling efficient exploration of materials design spaces.
\end{itemize}

\section{Method}

\begin{figure*}[p]
   \includegraphics[width=\textwidth]{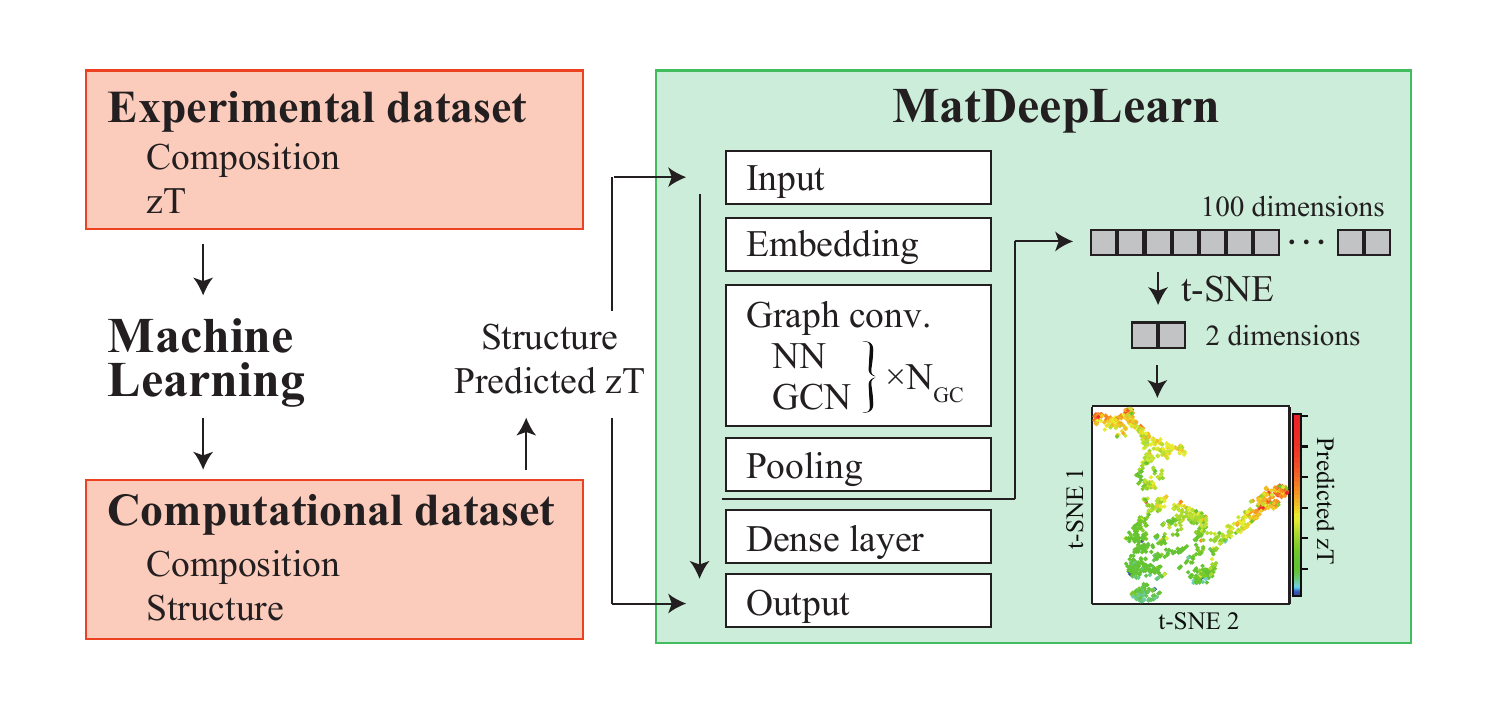}
   \caption{Schematic representation of the data flow and data analysis processes employed in this study.}
    \label{fig:dataflow}
\end{figure*}

We use a dataset extracted in our previous study~\cite{Jia2024-qq}, composed of the predicted experimental $zT$ values with the corresponding materials structures.
As briefly summarized in Appendix A, our approach consists of the following steps: preprocessing experimental data, training a machine learning model, and applying the trained model to predict experimental data for the compositions registered in the computational data. 
As a result, we obtained a dataset of the predicted experimental values of $zT$ with the structural information of the corresponding materials.

MatDeepLearn (MDL), developed by Fung $et$ $al$., provides an efficient Python-based environment for developing material property models~\cite{Fung2021-mg}. 
MDL supports the implementation of the graph-based representation of materials structures, deep learning modelling, and the construction of material maps via dimensional reduction using the t-SNE algorithm~\cite{Fung2021-mg}.
The open-source nature and extensibility of MDL make it valuable for researchers implementing graph-based materials property predictions with deep-learning based architectures~\cite{Fung2021-mg}.

The machine learning model defined in MDL is composed of input, embedding, graph convolution, pooling, dense, and output layers (Figure~\ref{fig:dataflow}).
The input layer primarily extracts basic structural information, such as atomic positions, types, and bond distances, using the Atomic Simulation Environment (ASE) framework.
The entire model is trained using material structures as input and the corresponding predicted $zT$ values as output.
After the model is trained, materials maps are constructed using the t-SNE architecture with the data extracted from the first dense layer (Figure~\ref{fig:dataflow}).

MDL supports various models for graph-based representation of materials structures, including CGCNN~\cite{Xie2018-rf}, Message Passing Neural Networks (MPNN)~\cite{Gilmer2017-ox}, MatErials Graph Network (MEGNet)~\cite{Chen2019-wm}, SchNet~\cite{Schutt2017-ca}, and Graph Convolutional Networks (GCN). 
In MDL, the graph-based model is defined by the Graph Convolutional (GC) layer, which is characterized by GC type, GC dimensions, and the repetition number of GC blocks (N$_{\text{GC}}$).
In this study, N$_{\text{GC}}$ was set to 4, which is the default hyperparameter of MDL~\cite{Fung2021-mg}, unless otherwise specified. 


\section{Results and discussion}

\subsection{Material map generated using MPNN architecture}

\begin{figure}
   \includegraphics[width=\columnwidth]{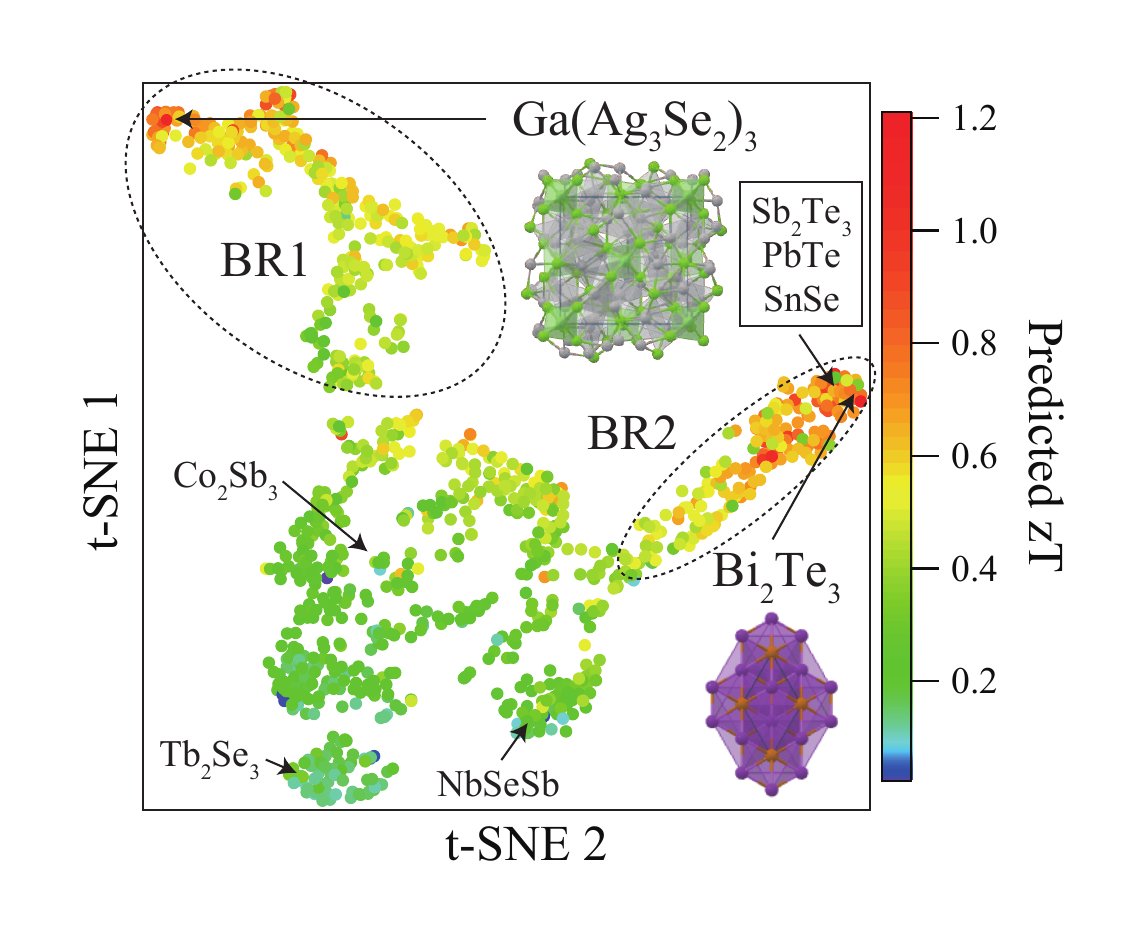}
   \caption{A materials map generated by MDL using MPNN architecture for graph-based modeling of materials properties. 
The color of each data point represents the predicted-experimental $zT$ values.
The position of each data point reflects the structural properties of materials extracted by the model trained by MDL.
Two branches spreading laterally are labeled BR1 and BR2 in the map. 
The structures of Bi$_2$Te$_3$ and Ga(Ag$_{3}$Se$_{2}$)$_{3}$ are demonstrated to illustrate the different complexities among the materials included in the two branches.}
    \label{fig:main}
\end{figure}

Let us first demonstrate in Figure~\ref{fig:main} the material map generated using the MPNN architecture~\cite{Gilmer2017-ox}.
The $y$ and $x$ axes represent the first and second t-SNE components, while the color of each point represents the predicted experimental $zT$ value.
The default hyperparameters of MDL~\cite{Fung2021-mg} were used in the model training.
In the figure, we find the smooth color gradients reflecting meaningful relationships between predicted $zT$ values and structural features.
A clear trend is observed, with lower $zT$ values concentrate in the lower region and higher $zT$ values appear in the right- and left-upper regions.
Moreover, the fine structures visible in the map suggest that the model effectively captures structural features of materials.
We also find two distinct branches, labeled BR1 and BR2 in Figure~\ref{fig:main}, extended laterally in the upper portion, featuring significant clusters of high $zT$ values at their endpoints.

\begin{figure}
    \centering
    \includegraphics[width=\columnwidth]{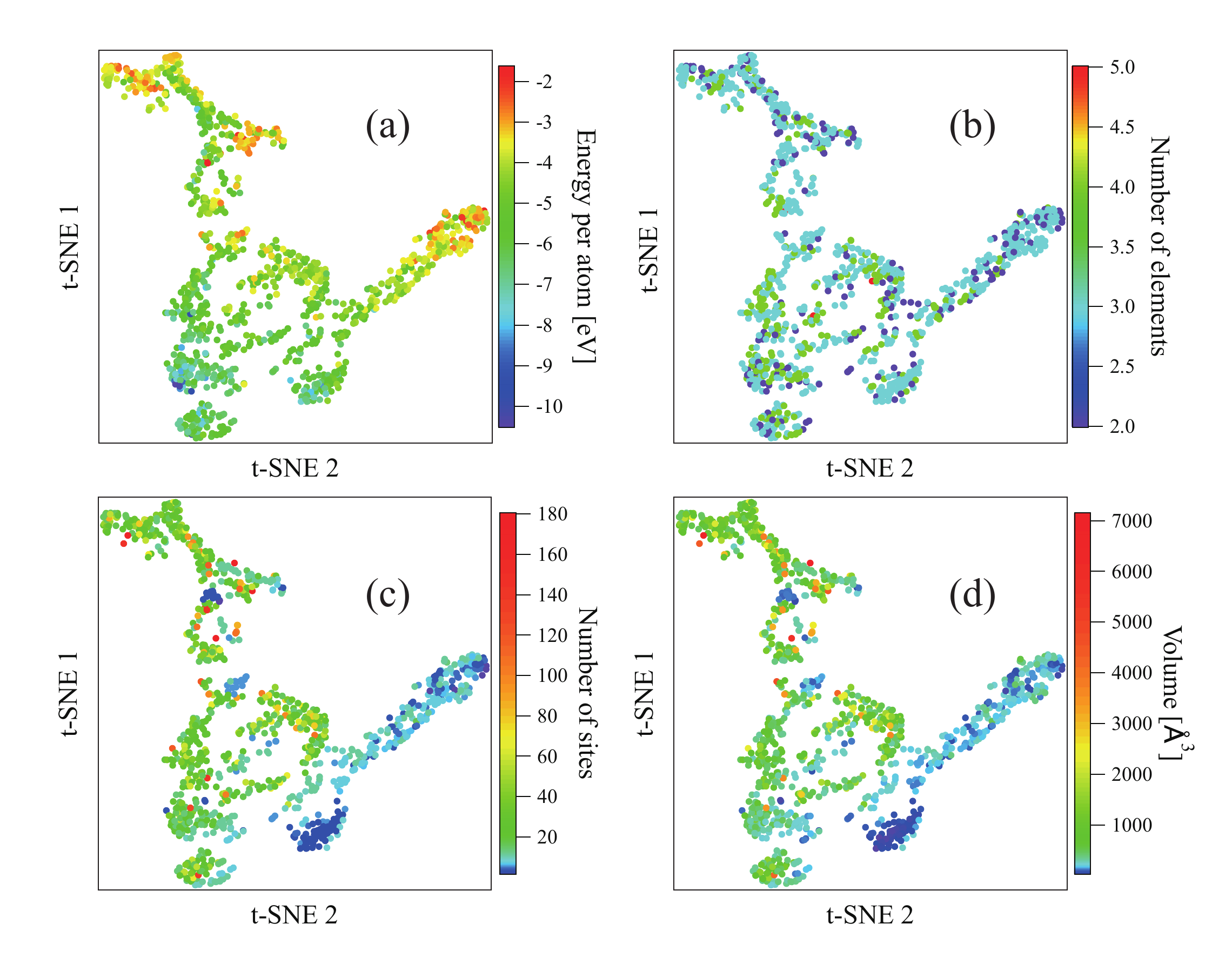}
    \caption{Material maps generated from the same data used in Figure~\ref{fig:main}, but color-coded by (a) energy per atom, (b) number of elements, (c) number of sites, and (d) volume. 
}
    \label{fig:dist}
\end{figure}

The computational dataset obtained from the Materials Project encompasses diverse materials data.
To improve the interpretability of the obtained maps, we created materials maps colored by other properties.
Figure~\ref{fig:dist} shows the maps colored by (a) energy per atom, (b) number of elements, (c) number of sites, and (d) volume.
The following analysis using these rich computational datasets with materials maps provides detailed insights into the obtained maps.

In Figure~\ref{fig:main}, colored by predicted $zT$, we find the lower values in the bottom left side while high values in the top right and top left sides. 
Similar trends are found in Figure~\ref{fig:dist}(a), colored by the energy per atom, representing the close relationships between the corresponding parameters.
This trend is supported by the high correlation coefficient of 0.66, obtained through the correlation analysis described in Appendix B.
In contrast, Figure~\ref{fig:dist}(b), colored by the number of elements, shows the uniform distribution of the data points. This trend reveals that the structures shown in the map do not represent the complexities in the material composition.

In the maps, we find two branches spreading laterally, labeled as BR1 and BR2 in Figure~\ref{fig:main}.
The difference between these two branches is investigated by the maps colored by the number of atomic sites and the volume shown in Figures~\ref{fig:dist}(c) and~\ref{fig:dist}(d), respectively.
In these figures, we observe high and low values for BR1 and BR2, respectively.
These results suggest that the difference between these two branches arises from variations in structural complexities.
As is schematically represented in Figure~\ref{fig:main}, Ga(Ag$_3$Se$_2$)$_3$, representing BR1, exhibits a highly complex structure, whereas Bi$_2$Te$_3$, representing BR2, displays an extremely simple structure. 

\subsection{Distribution of materials in material map}

\begin{figure}
\centering
\includegraphics[height=0.9\textheight, keepaspectratio=true]{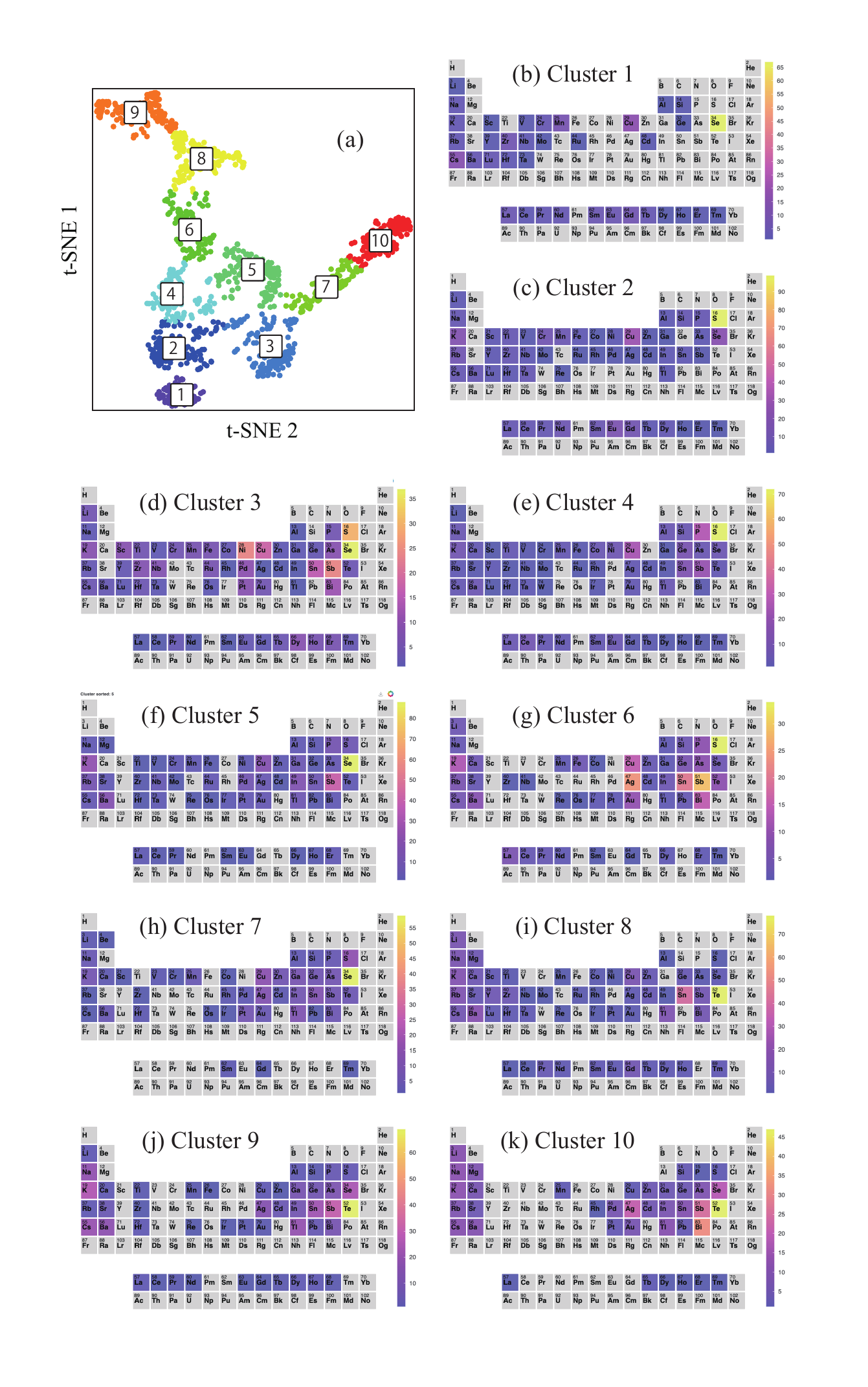}
\caption{(a) A material map colored by cluster numbers obtained via $k$-means clustering with $k = 10$. (b-k) The elemental compositional analysis results for materials included in each cluster.}
\label{fig:cluster}
\end{figure}

To investigate the material distribution in the map, we applied $k$-means clustering analysis to the map with $k$=10 clusters as shown in Figure~\ref{fig:cluster}(a). The label for each cluster was assigned by sorting based on the average of predicted $zT$ values, allowing us to identify regions of high-performance materials. To characterize the materials included in each cluster, we counted the frequency of elements in the compositions of materials within each cluster. The elements with less than 10$\%$ of the compositional ratio were excluded to focus on elements that make major contributions to the material properties. The results for each cluster are plotted using the Periodic Trend Plotter (https://github.com/Andrew-S-Rosen/periodic\_trends) in Figure~\ref{fig:cluster}(b)-(k). In each cluster, we find distinct characteristic distributions of S, Se, and Te atoms. Clusters 1–7 containing materials with low $zT$ values show a high occurrence frequency of S and Se. On the other hand, clusters 8–10 containing materials with high $zT$ values exhibit a higher frequency of Te occurrences. To investigate the characteristics of the two branches, labeled as BR1 and BR2 in Figure~\ref{fig:main}, we compare the elemental distribution in clusters 9 and 10, which are located at the edge of each branch. We find high similarities between these two plots, such as the high frequencies of Te, Sb, Ti, and Ag atoms. Namely, the materials in clusters 9 and 10 have similar compositions, but are separated in the map due to differences in structural complexity.

\subsection{Graph-based model dependency of material map}

\begin{figure}
   \includegraphics[width=\columnwidth]{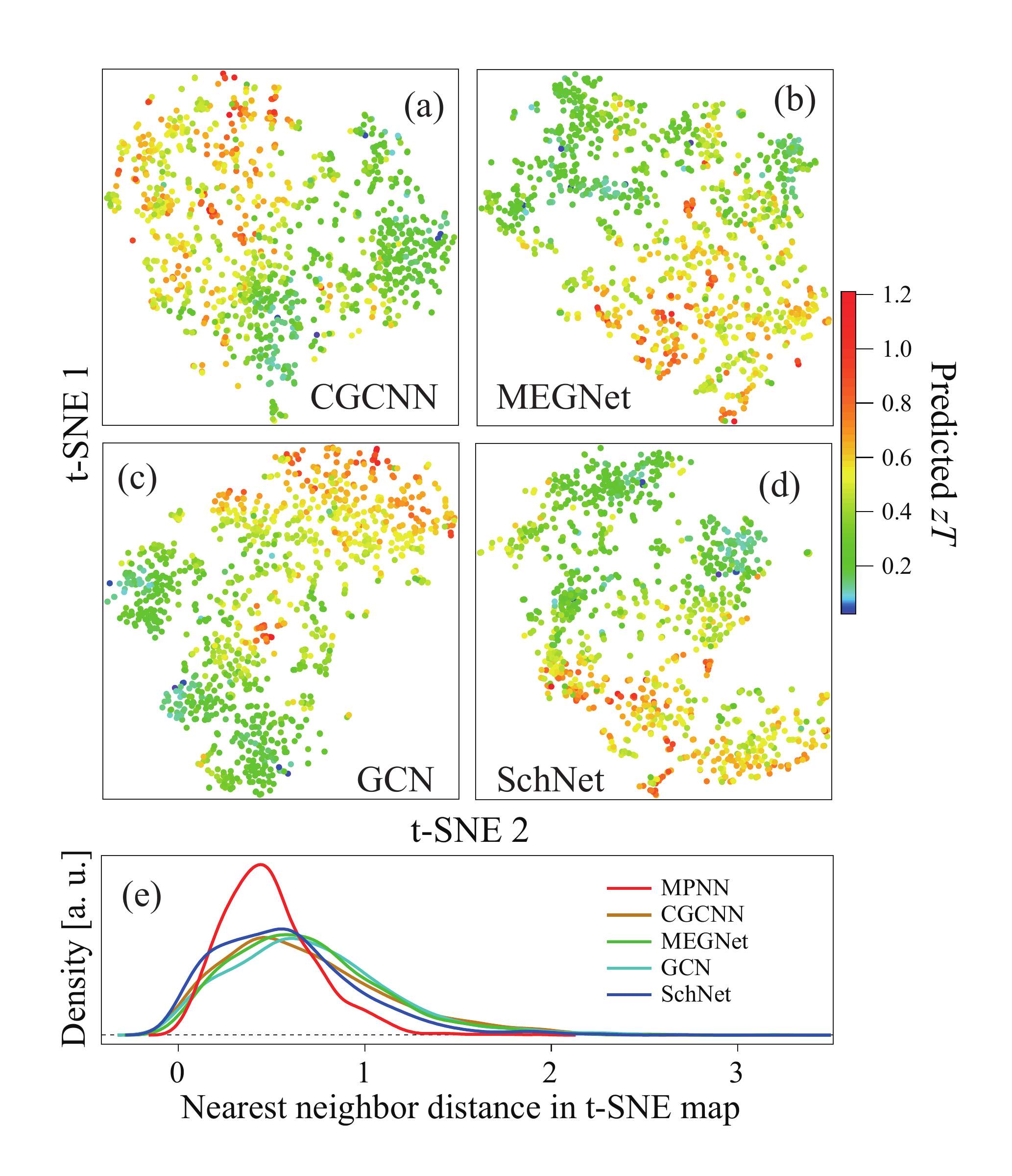}
   \caption{Comparison of material maps generated by MDL using different graph-based architectures: (a) CGCNN, (b) MEGNet, (c) GCN, and (d) SchNet. (e)  The distributions of data points in the maps shown in (a)-(d) are compared by the distribution analysis of NND of each data point by KDE.}
    \label{fig:model}
\end{figure}

We have discussed material maps generated by MDL using the MPNN architecture, while MDL also supports other architectures, such as CGCNN~\cite{Xie2018-rf}, MEGNet~\cite{Chen2019-wm}, GCN, and SchNet~\cite{Gilmer2017-ox}.
In Figures~\ref{fig:model}(a)-\ref{fig:model}(d), we compare the material maps generated by each of these models and find that the maps generated by the other models lack the clear structures observed in the map generated by MPNN. 
To quantify the structures of data points in the map, we evaluate the distribution of the nearest neighbor distances (NND) of each data point by Kernel Density Estimation (KDE)~\cite{Parzen1962-lg}.
KDE is a non-parametric method for estimating smooth probability density functions from data points, and thus offering a smoother and less biased representation than histograms.
In Figure~\ref{fig:model}(e), we find that the KDE spectrum for the MPNN-generated map displays a higher density in the low NND region compared to the spectra for other models.
This trend indicates the superiority of MPNN as an effective tool for providing more structurally organized and continuous mappings compared to the other graph-based architectures.

The model dependence for materials map constructions is also discussed in Ref.~\onlinecite{Fung2021-mg}, which is the first report of MDL. This study compares materials maps generated with different graph-based architectures for the structural feature extractions; however, the differences in the obtained maps were not as significant compared to the present case. We attribute the difference in these two studies to the different data screening. Although both studies use the same database, the Materials Project, our study applied filters based on material stability, band gap, and constituent elements. This screening resulted in the selection of similar materials, which facilitated feature extraction by the machine learning, leading to material maps with clearer structures.



\subsection{Relationship between material property prediction and material map}

The machine learning model defined in MDL predicts materials properties and generates material maps as part of a continuous process.
(Note that this prediction by MDL is different from the prediction of the experimental $zT$ values, described in Ref.~\onlinecite{Jia2024-qq}.) 
Intuitively, we can expect that the creation of a high-quality material map is accompanied by the effective extraction of materials features, and thus leads to enhanced accuracy in material property predictions.
This logic motivates us to evaluate the model performance for the material property prediction by three standard metrics: the coefficient of determination (R$^2$), mean squared error (MSE), and mean absolute error (MAE) with the data divided into training (80$\%$), testing (15$\%$), and validation (5$\%$) subsets.
The evaluation results with the validation dataset are summarized in Table ~\ref{tab:model_accuracy}.
While MPNN produces the most structured and interpretable maps, its prediction accuracy (R$^2$ = 0.610) is the lowest among all the models.
This unexpected result means that the clear structural organization displayed in the MPNN-derived maps does not necessarily translate to higher predictive performance of the materials properties.

\begin{table}[h]
\centering
\begin{tabular}{lccc}
\hline
Model & MSE & MAE & R$^2$ \\
\hline
MPNN & 0.014594 & 0.083548 & 0.610463 \\
CGCNN & 0.011580 & 0.076089 & 0.690913 \\
MEGNet & 0.009895 & 0.070994 & 0.735893 \\
GCN & 0.007246 & 0.057658 & 0.806579 \\
SchNet & 0.013064 & 0.080413 & 0.651301 \\
\hline
\end{tabular}
\caption{Material property prediction performance for various graph-based architectures. MSE (Mean Squared Error), MAE (Mean Absolute Error), and R$^2$ (coefficient of determination) were used as evaluation metrics.}
\label{tab:model_accuracy}
\end{table}

\subsection{Contribution of MPNN architecture for materials map construction}

\begin{figure}
    \centering
    \includegraphics[width=\columnwidth]{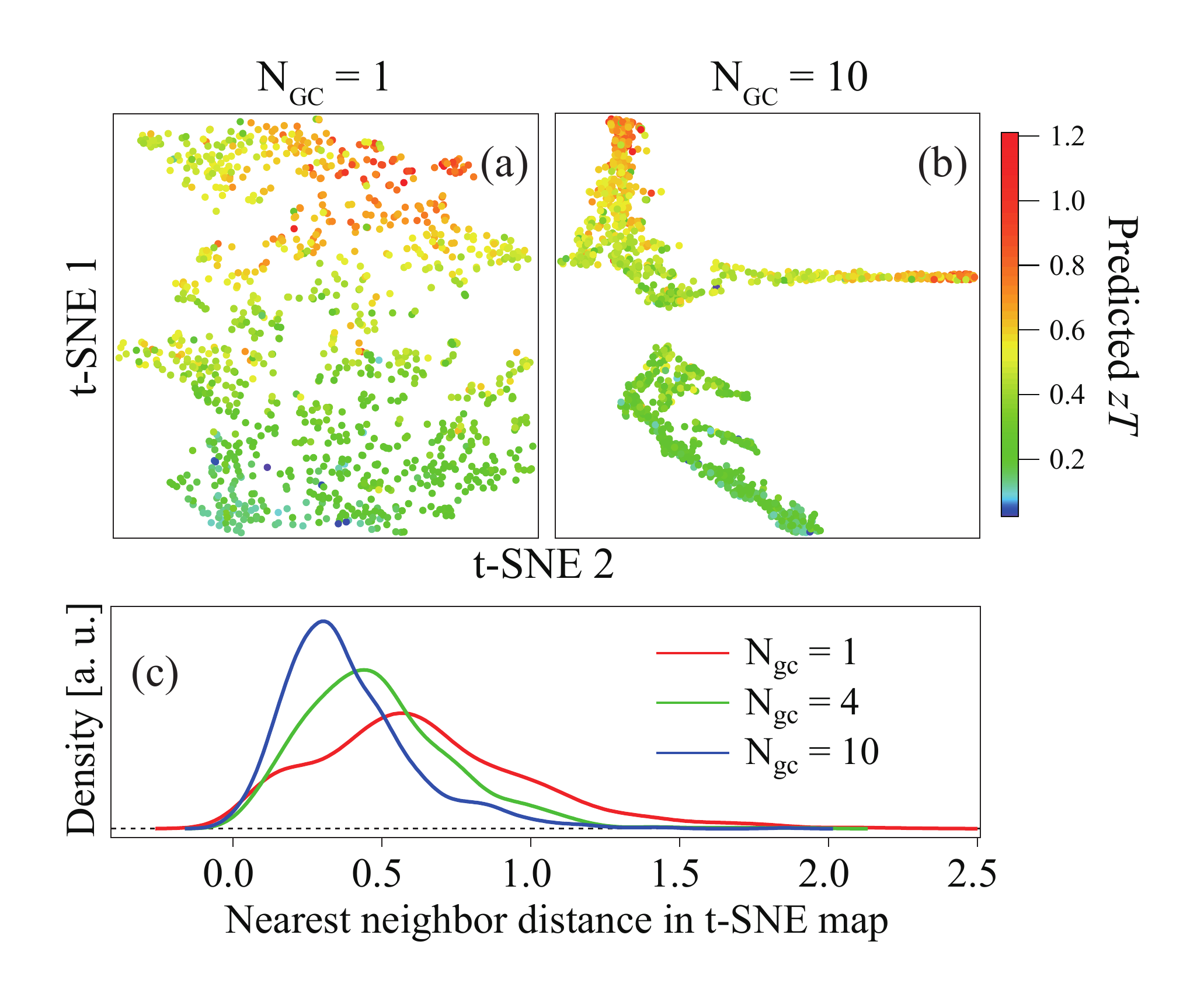} 
    \caption{Comparison of the material maps generated by MDL with (a) one GC layer (N$_{\text{GC}}$ = 1), and (b) ten GC layers (N$_{\text{GC}}$ = 10).  (c) The structures in the maps for N$_{\text{GC}}$ = 1, 4, and 10 are compared by the distribution analysis of NND for each data point by KDE.} 
    \label{fig:gc}
\end{figure}

As described above, MDL with MPNN architecture, implemented in the GC layer, generates well-structured materials maps.
To investigate the role of the MPNN architecture in the materials map construction, we compare maps generated with different repetition number of GC block (N$_{\text{GC}}$ = 1 and 10) in Figure~\ref{fig:gc}.
We find that the increase in N$_{\text{GC}}$ leads to tighter clustering of data points.
This trend is again evaluated by the distribution analysis of NND of each data point by KDE.
The KDE spectra obtained with N$_{\text{GC}}$ = 1, 4 (default value for MDL), and 10 are shown in Figure~\ref{fig:gc}(c).
We find that the peak intensity increases and the peak position shifts to the lower NND region with increasing N$_{\text{GC}}$.
These results indicate the enhanced feature learning by the GC layer. 

\begin{figure}[p]
    \centering
    \includegraphics[height=0.8\textheight]{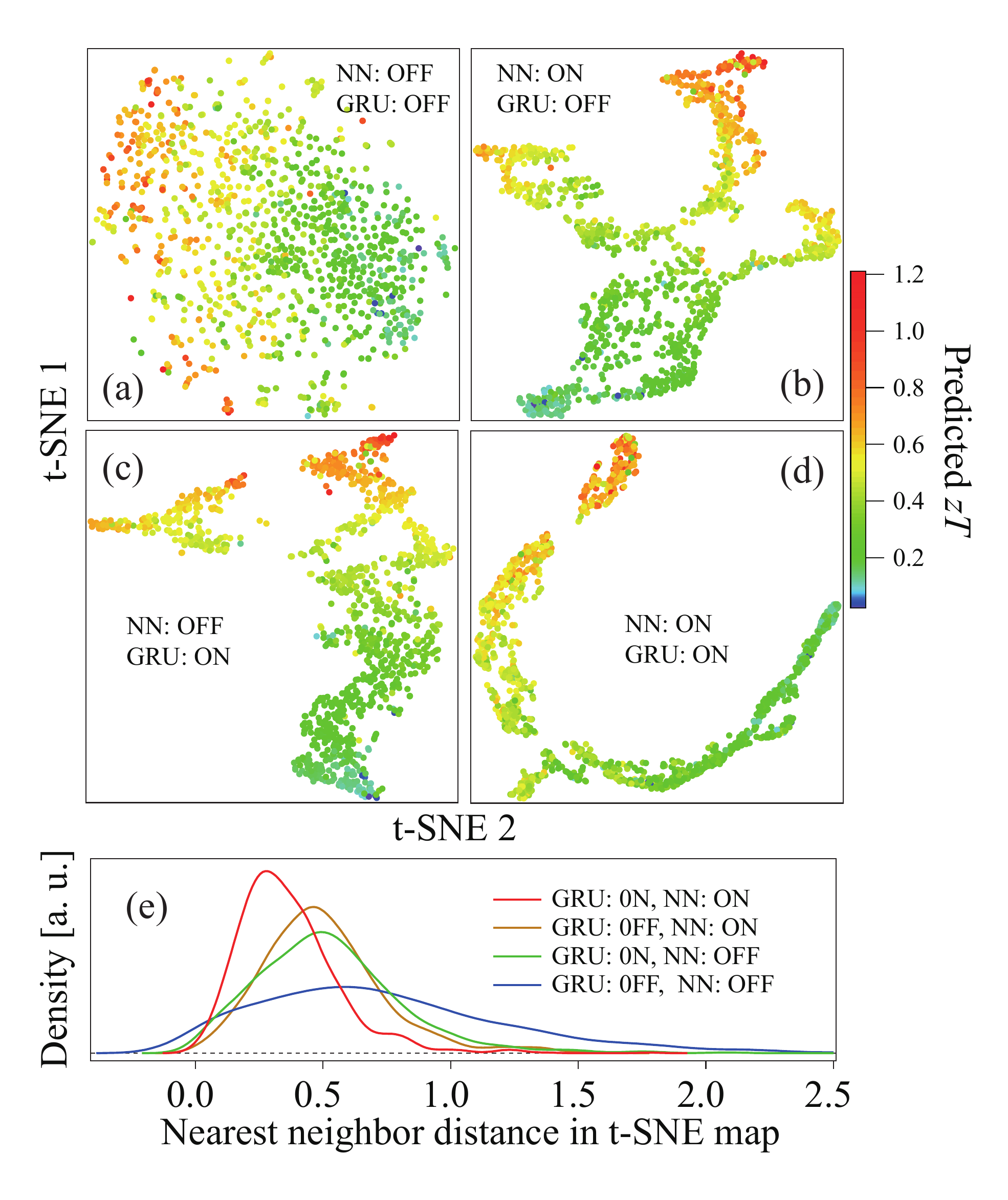} 
    \caption{Materials maps generated by MDL with different training configurations in the GC layer, composed of neural networks (NN) and gated recurrent units (GRU). To emphasize the contribution of the GC layers, N$_{\text{GC}}$ = 10 was selected.  The training for each unit is separately controlled as (a) NN: OFF, GRU: OFF, (b) NN: ON, GRU: OFF, (c) NN: OFF, GRU: ON, and (d) NN: ON, GRU: ON. Each map illustrates how the configurations of the GC layer influence the distribution of data points in the materials map. The structures in the maps shown in (a)-(d) are compared by the distribution analysis of NND of each data point by KDE.}
    \label{fig_NN_GRU} 
\end{figure}

In MDL, MPNN architecture is implemented through the NN and GRU blocks in the GC layer~\cite{Fung2021-mg}.
The NN block enhances the model's representational capacity, while the GRU block improves learning efficiency through memory mechanisms.
To investigate the contributions of each block, we selectively controlled the training of each block, turning them ON or OFF, by modifying the MDL code. 
N$_{\text{GC}}$ was again set to 10 to enhance the contribution of each block.
Figures~\ref{fig_NN_GRU}(a)-(d) illustrate the effects of disabling learning in either the NN or GRU blocks. 
When learning is disabled in either block, the points spread more diffusely; when both are disabled, the points distribute uniformly across the map.
These trends are again clearly captured by KDE as shown in Figure~\ref{fig_NN_GRU}(e).
By turning ON the training by either or both of the NN and GRU blocks, the peak intensity increases and the peak position shifts toward the lower NND region.
These results convince us that the learning by the GC layer, composed of the NN and GRU blocks, plays a crucial role in the structural organization in the materials maps.

MPNN is a graph-based architecture developed relatively early in the field and is characterized by high learnability and versatility~\cite{Gilmer2017-ox}.
The dominant contribution by the MPNN architecture for the structural formation in the obtained materials map is evidenced as discussed above.
However, due to the black-box nature of deep learning-based modeling, it is unclear why the MPNN-generated materials map can capture materials properties so efficiently.
Further investigation into the reasons behind these discrepancies could yield valuable insights into how structural interpretability and predictive power might be balanced in future model designs.

\subsection{Finding proper materials with material map}

Finding optimal materials is extremely important but is always challenging due to the enormous variety of candidate materials.
High-throughput screening of materials using computational databases and machine learning is a powerful tool that can suggest promising materials.
However, experimental researchers of materials frequently face difficulties in developing suggested new materials, due to limited access to raw materials, equipment, facilities, and experimental expertise. 
Therefore, it is more important to explore promising materials among those that experimental researchers can realistically fabricate.
Generally speaking, materials with similar structures are likely to be synthesized and characterized using similar methods. 
For example, researchers working on high-entropy materials~\cite{Ma2023-ms}, which have significantly complex compositions and structures, require different knowledge and experimental equipment compared to those investigating ordinary materials.
The materials maps that automatically distinguish materials by their structural characteristics, especially complexity, can support experimental researchers in finding their next target materials. 

\begin{figure}
\includegraphics[width=\columnwidth]{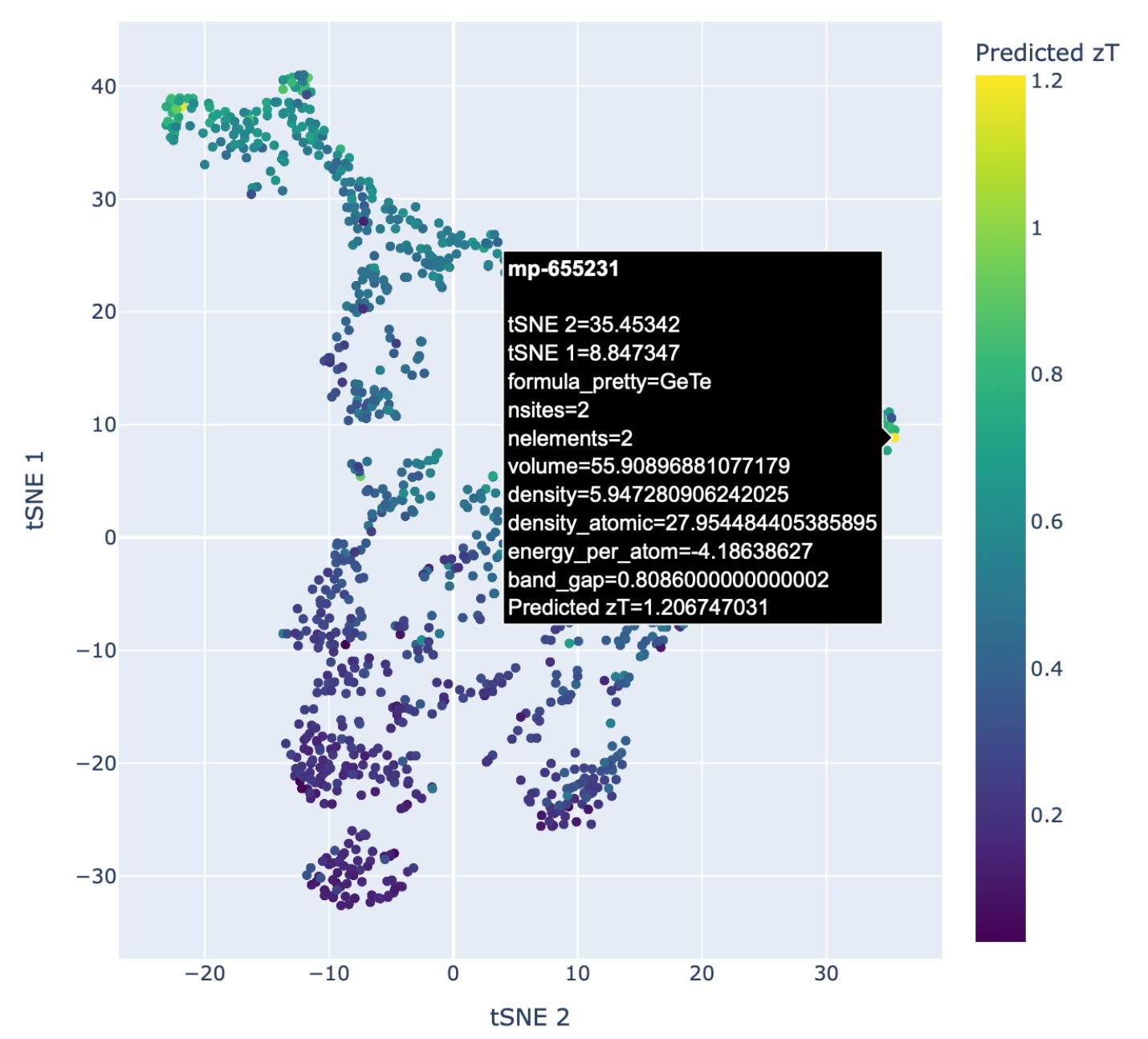}
\caption{Interactive graph of materials properties made by Plotly library with the data used in Figure~\ref{fig:main}.}
\label{fig:interactive_graph}
\end{figure}

An interactive graph is a powerful tool for visualizing complex and multi-dimensional data. Using Python libraries such as Plotly, interactive graphs can be easily created, and when saved as HTML files, the data can be accessed at any time. In Figure~\ref{fig:interactive_graph}, we show the interactive graph of the same data used in Figure~\ref{fig:main}. This map can support targeting specific material properties by referencing aspects such as composition, data names, and other material properties, enabling multifaceted and detailed data exploration. The corresponding HTML file is included in the supplementary materials.

\subsection{Prospects for future studies}

Data-driven materials development has a great potential to revolutionize research and development in various domains, such as inorganic, organic, nano- and pharmaceutical materials.
Machine learning modeling with the graph-based representation of materials structures may answer the challenge to construct a truly universal materials model that applies to all materials~\cite{Shoghi2023-oh}.
This study focused on the thermoelectric properties of materials containing six selected elements (Sb, Te, Sn, Se, Bi, S), which demonstrated predictive capability by machine learning models trained by experimental data of $zT$ values~\cite{Jia2024-qq}. 
Extending this model to other physical properties and a wider range of elements is left for future studies.

The materials maps demonstrated in this study reflect structural complexity of materials since MDL limits the input data to structural information.
By selecting input features, we can tailor the map to meet specific objectives or explore different aspects of materials behavior.
A key avenue for future work involves integrating additional characteristic variables, such as magnetic, chemical, or even topological properties of materials, to create a more comprehensive materials map.

The creation of materials maps requires substantial computational resources, especially for tasks such as dimensionality reduction by the t-SNE architecture and the deep-learning modeling defined in MDL. 
In this study, we employed the model and the hyperparameters given in MDL due to computational constraints. 
For wider and better applicability, fine-tuning of the model may be important.
Recent advancements in high-performance computing and GPU/TPU clusters will help mitigate these computational challenges, enabling broader adoption of such methods.
Note that the memory usage for computation increases dramatically with N$_{\text{GC}}$ especially when large datasets are used for the data analysis~\cite{Omee2022-tk}. This is demonstrated by the study using DeeperGATGNN, which allows modeling with large N$_{\text{GC}}$~\cite{Omee2022-tk}. Therefore, the use of MDL with MPNN may be limited to cases where dataset size is small or when computing resources with sufficient memory size are available.

In our study, we integrate computational and experimental data by predicting experimental $zT$ values from the compositions stored in the computational database~\cite{Jia2024-qq}.
However, another approach—predicting stable structure from composition—is also conceivable. 
One should select the method based on the accuracies of the machine learning models, opting for the approach that achieves higher predictive accuracy.
In this study, we employed our approach due to the high prediction accuracy of our model ($R^2 \sim$ 0.9), predicting experimental $zT$ values from compositions~\cite{Jia2024-qq}, and the technical difficulty in the prediction of the stable structure by composition~\cite{Gusev2023-bz}. 


\section{Conclusion}

We demonstrated the construction and evaluation of material maps generated by the MatDeepLearn (MDL) framework and the integrated datasets of computational and experimental studies. The materials map generated by the MDL framework with MPNN architecture clearly represents the structural complexities, exhibiting smooth color gradients reflecting the predicted experimental $zT$ values. The structures in the maps are statistically evaluated using the distribution analysis of nearest neighbor distances (NND) for each data point through Kernel Density Estimation (KDE). Systematic analysis of material maps across different types, structures, and training configurations of graph-based models reveals the superior performance of MPNN for generating well-structured material maps that represent the structural similarities of materials.

Materials maps generated using integrated experimental and computational datasets hold great potential to bridge the gap between these approaches. Since materials with similar structures tend to be synthesized and evaluated with similar methods experimentally, material maps reflecting the similarity of material structures can support the targeting of promising materials by experimental researchers, and thus contribute to efficient and enhanced materials discovery.

\section{Author contributions}
Li H and Takaaki T supervised the project; Yusuke H conceived the idea; Jia X performed the data preprocessing; Yusuke H performed data analysis and wrote the paper; all authors
discussed the results; all listed authors agree to all manuscript contents, the author list and its order, and the author contribution statements.

\section{Conflict of interest}
The authors declare that they have no conflict of interest.

\section{Data availability}
The corresponding raw data files as well as further data that support the findings of
this work are available from the corresponding authors upon reasonable request.

\section{Code availability}
The codes used for data analysis and simulated results are available from the corresponding authors upon reasonable request.

\section{Acknowledgments}
We thank the AIMR ATP project in Tohoku University and Prof. Yong P. Chen’s group for computational support.

\bibliographystyle{tfnlm} 

\bibliography{bibtex} 

\appendix


\section{Data preparation and machine learning}

\subsection{Cleaning of $zT$ data in StarryData2}
We use the datasets obtained from the StarryData2 (SD2), an open database for experimental material properties.
The initial dataset from SD2 contained various sources of noise and errors, including those from original papers, data mining processes, inadvertent inclusion of calculation data, and extreme experimental conditions. To ensure data quality, we implemented a rigorous cleaning process:

\begin{itemize}
\item Removed compositions with typographical errors
\item Excluded entries with titles containing keywords indicative of calculations
\item Filtered out $zT$ values less than 0 or greater than 3.1
\item Omitted studies published before the year 2000
\end{itemize}

After this cleaning process, 8,541 compositions were selected from SD2.

\subsection{Machine learning modeling of experimental dataset}

For feature extraction, we utilized Python libraries from matminer, specifically the $elementproperty.from\_preset(magpie)$ function. The machine learning model was developed using the following approach:

\begin{itemize}
\item Algorithm: Gradient Boosting Decision Tree
\item Validation: 10-fold cross-validation
\item Data split: 80$\%$ training, 20$\%$ testing
\end{itemize}

This process resulted in a machine learning model with an R$^2$ value of 0.85. 
Additional data cleaning, which excludes the data showing poor prediction accuracy, further improves the model, leading to $R^2 = 0.90$.
Through the composition analysis of the selected materials, we selected Sb, Te, Sn, Se, Bi, and S as key elements for further investigation.

\subsection{Machine learning modeling of calculation dataset}

To expand our analysis to computational data, we applied our experimental data-trained model to materials from the Materials Project database:

\begin{itemize}
\item Selected compositions containing Sb, Te, Sn, Se, Bi, and S
\item Filtered for band gap energy above 1.0 eV
\item Further filtered for energy hull above 0 eV
\end{itemize}

As a result, 1,114 compositions were selected. By applying our experimental $zT$ prediction model to this refined set of computational data, we were able to predict experimental $zT$ values for these materials. This approach led to the identification of GeTe$_5$As$_2$ and Ge$_3$(Te$_3$As)$_2$ as potentially high-zT materials.

\section{Correlation analysis of parameters}

The correlation between the predicted $zT$ values and the parameters stored in the computational datasets (Materials Project) is analyzed using the '.corr()' function in the pandas framework for Python. The correlation coefficient $r$ is defined as follows:

\begin{subequations}
\renewcommand{\theequation}{1\alph{equation}}
\begin{align}
r &= \frac{\sum_{i=1}^{n} (x_i - \bar{x})(y_i - \bar{y})}{\sqrt{\sum_{i=1}^{n} (x_i - \bar{x})^2}\sqrt{\sum_{i=1}^{n} (y_i - \bar{y})^2}}, \\
\text{, where} \nonumber \\
\quad \bar{x} &= \frac{1}{n}\sum_{i=1}^{n} x_i, \\
\bar{y} &= \frac{1}{n}\sum_{i=1}^{n} y_i.
\end{align}
\end{subequations}

The example of $r$ calculated between the predicted $zT$ values and the selected properties stored in the Materials Project are summarized in the following.

\begin{table}[h]
\centering
\begin{tabular}{lc}
\hline
Parameter name & Correlation coefficient\\
\hline
Energy per atom & 0.65664\\
Volume & 0.05841\\
Density & 0.02604\\
Number of sites & -0.01809\\
Number of elements & -0.16573\\
\hline
\end{tabular}
\caption{Correlation coefficient between predicted $zT$ and the parameters stored in the Materials Projects.}
\label{tab:model_comparison}
\end{table}

\section{Dimensionality reduction for material map construction}

In this study, we generate material maps by dimensional reduction using the t-SNE algorithm, which is implemented using the scikit-learn framework, as in MDL.
The default hyperparameters in MDL, which are a perplexity of 50 and a learning rate of 300, were used for the dimensional reduction by t-SNE.

\section{Hardware and software environment}
Calculations for deep learning-based modeling in this study were performed on a workstation equipped with three NVIDIA RTX 3090 GPUs, 128 GB RAM, and an Intel Core i9-10900X CPU @ 3.70GHz. The software used included Python 3.10, TensorFlow 2.10, and the MDL framework. 

\end{document}